# Accurate Measurements of Free Flight Drag Coefficients with Amateur Doppler Radar

Elya Courtney, Collin Morris, and Michael Courtney
Michael_Courtney@alum.mit.edu

**Abstract:** In earlier papers, techniques have been described using optical chronographs to determine free flight drag coefficients with an accuracy of 1-2%, accomplished by measuring near and far velocities of projectiles in flight over a known distance. Until recently, Doppler radar has been prohibitively expensive for many users. This paper reports results of exploring potential applications and accuracy using a recently available, inexpensive (< $600 US) amateur Doppler radar system to determine drag coefficients for projectiles of various sizes (4.4 mm to 9 mm diameter) and speeds (M0.3 to M3.0). In many cases, drag coefficients can be determined with an accuracy of 1% or better if signal-to-noise ratio is sufficient and projectiles vary little between trials. It is also straightforward to design experiments for determining drag over a wide range of velocities. Experimental approaches and limitations are described. Overall, the amateur radar system shows greater accuracy, ease of use, and simplicity compared with optical chronographs. Doppler radar has advantages of working well with less accurate projectiles without putting equipment at risk of projectile impact downrange. The system can also detect phenomena that optical chronographs cannot, such as projectile instability resulting in tumbling in flight. This technology may be useful in introductory physics labs, aerodynamics labs, and for accurately determining drag and ballistic coefficients of projectiles used in military, law enforcement, and sporting applications. The most significant limitations are reduced signal-to-noise with smaller projectiles (< 5 mm diameter) and inability to detect projectiles more than 100 m down range.

**Keywords**: *drag coefficient, ballistic coefficient, free flight, wind tunnel, optical chronograph, Doppler radar, physics labs*

## Introduction

Drag coefficients and/or the equivalent ballistic coefficients are used to predict projectile trajectories, wind drift, and kinetic energy retained downrange. They are often of academic interest in undergraduate laboratories as students learn more accurate methods to account for air resistance. However, there are few simple and inexpensive experimental approaches to measuring drag coefficients with accuracy of better than 5-10%.

Ballistic coefficients provided by the manufacturers are often inaccurate or apply only under ideal conditions. (Courtney and Courtney, 2007; Litz, 2009a; Halloran et al., 2012; Bohnenkamp et al., 2012). Air drag may also depend on the bore from which a projectile is launched (Bohnenkamp et al., 2011). Further, drag coefficients can depend on gyroscopic stability and velocity of a projectile (Courtney and Miller, 2012a, 2012b; McDonald and Algren, 2003). An accurate measurement technique is needed to measure drag coefficients with an accuracy better than 5-10%.

For a given instrumental uncertainty, the most accurate approach to measuring drag coefficients is by measuring a near and a far velocity over a specified distance (Courtney et al., 2015; Bailey and Hiatt, 1972). Simple calculations using classical mechanics and the definition of drag coefficients yield the experimental values. Ballistic calculators may be used to yield ballistic coefficients with a given near and far velocity (www.jbmballistics.com).

It has previously been shown that with adequate experimental care and calibration, drag coefficients can be determined to an accuracy of 1-2% using optical chronographs designed for the sporting market (Courtney et al., 2015). While accurate and inexpensive, this method has disadvantages, including the requirements to place the projectile in a small area passing over the chronographs, to carefully measure the chronograph separation, to calibrate the chronographs daily, and to pay constant attention to other details like skyscreen angles.

A new amateur Doppler radar based chronograph has recently come onto the sporting market (www.mylabradar.com). An early review suggested the feature of recording velocity as a function of distance might prove useful for accurately determining ballistic coefficients (https://www.shootingsoftware.com/doppler.htm).

The purpose of this paper is to explore accuracy potential and ease of use in using amateur ballistic radar to experimentally determine projectile drag coefficients.

The method section describes basic use of the radar system, but also focuses on two main analysis options for determining drag and ballistic coefficients: 1) using the velocity at two ranges as

# Accurate Measurements of Free Flight Drag Coefficients with Amateur Doppler Radar

determined by the unit itself and 2) using the raw velocity vs. distance data determined at a number of ranges. The second approach has the advantage of allowing an uncertainty estimate to be computed for each trial or each shot; whereas, estimating the uncertainty from the first method requires computing the standard error of the mean (SEM) of a number of shots.

The results sections present drag coefficient results and estimated uncertainties from a number of test cases. 30 caliber supersonic rifle bullet results are presented first, followed by 22 caliber supersonic rifle bullets. After that results are presented for transonic and subsonic pistol bullets followed by results for subsonic 17 caliber bbs.

The discussion summarizes accuracy in the test cases and provides an assessment of strengths and weaknesses. Potential uses of the radar are suggested, ranging from professional work to educational laboratories and sporting applications.

Because ballistic coefficients are of greater interest in many military, law enforcement, and sporting applications, ballistic coefficient results are presented in the appendix.

**Methods**
Experimentally determined drag coefficients require measurements of near and far velocities, separation distance between velocity measurements, air density, cross sectional area of the projectile, and projectile mass (Equation 4 Courtney et al., 2014). Cross sectional area is computed from projectile diameter measured with a dial caliper. Mass is determined with a laboratory scale. Air density is computed from directly measuring barometric pressure, temperature, and relative humidity with a Kestrel 4500 weather meter. As described in Courtney et al. (2015), the uncertainties in most inputs are less than 0.3%. The main limitation in accurate determination of drag coefficients in the absence of significant wind is accurate determination of the near and far velocities.

The LabRadar advertises an accuracy of 0.1% in velocity readings,[1] but the marketing claim is ambiguous regarding whether this applies to every reading at every distance, every muzzle velocity it determines, or is a typical expectation for muzzle velocity. After working with the unit and analyzing the raw Doppler velocity vs. distance data it uses to determine muzzle velocities, it seems clear that the 0.1% accuracy specification is only reasonable as a typical expected accuracy for muzzle velocity. This conclusion is clear, because downrange velocity readings often show significant noise, so a claim of 0.1% accuracy of the raw data would demonstrate many cases of the physical impossibility of the bullet gaining velocity as it moves downrange.

LabRadar has not documented exactly how the raw Doppler velocity vs. distance data is used to determine a muzzle velocity (V0) and velocity at 5 additional user-selected distances. It appears that a kind of regression is performed on the raw data, and the best fit model is used to determine V0 and velocities at the set distances, which are then displayed to the user and recorded in the Report file for that shot string.

The unit always extrapolates backward for V0 (since all the Doppler readings are at positive distances in front of the LabRadar). However, the internal firmware seems unwilling to extrapolate velocity readings at distances much larger than those for which it has Doppler velocity reading available. The raw Doppler velocity vs. distance readings for each shot are available in a Track (*.TRK) file for each shot recorded.

The Track files contain columns for time, velocity, distance and signal-to-noise ratio (SNR). It seems that data is recorded at fixed times out to a distance approximately 10 yards beyond the furthest distance the user sets to determine velocity and for which the SNR is above some threshold that seems close to 10.

These observations regarding instrument operation and data availability suggest two possibilities for computing drag coefficients: 1) use the displayed velocity readings at two distances as reported in the instrument panel and the Report file as the near and far velocities or 2) develop an analysis method based more directly on the raw data in the Track files. Both possibilities were pursued here to assess the uncertainties and accuracy potential of each, and weigh those against extra time and effort required to begin with the raw data.

*Instrument Alignment*
The LabRadar has internal instrument settings related to triggering, velocity range, internal file storage, and data acquisition rate. These are well

---

[1] Section 9 (Range) on p. 20 of the User Manual (v 1.1) discusses a number of factors that can influence both accuracy and range of a given instrument use. A careful reading of that section is recommended for users measuring drag coefficients.

# Accurate Measurements of Free Flight Drag Coefficients with Amateur Doppler Radar

described in the user's manual and relatively simple. In contrast, the most challenging aspect of operating the instrument is getting it physically aligned with the bullet path well enough for sufficiently accurate downrange velocity readings to compute meaningful drag coefficients.

The instrument's plastic case has a thin notch in the top, and the user manual instructs to point the notch at the target. However, in the unit we were using, this did not reliably produce downrange velocity readings for all calibers. Due to the relatively low power of the radar device, one needs fairly good overlap between the stronger part of the radar beam and the bullet path for sufficient signal-to-noise for reliable radar readings, especially for projectiles with smaller base diameters beyond 50 yards. We found a sturdy camera tripod necessary and useful for achieving and maintaining the careful alignment needed.

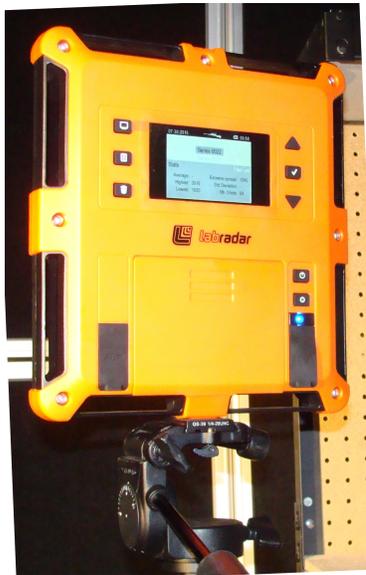

*Figure 1: Mounting the LabRadar instrument on a sturdy camera tripod was necessary for careful initial adjustment of and to maintain alignment.*

Over the first few data collection sessions, we found that the instrument was easiest to align with larger caliber projectiles (36 caliber or 9 mm). One could make coarse adjustment with these projectiles using feedback for which was the longest range for which the unit had a successful velocity reading displayed. (We had the unit configured to display velocities out to 70 yards, if available). Once the unit was reading successfully to 70 yards with 9mm projectiles, we kept the firing point and target the same, and more finely aligned the instrument to improve the readings with 30 caliber flat base projectiles.

One could then progress through 30 caliber boat tail projectiles (smaller base reflects less signal), 22 caliber flat base projectiles, etc. The alignment process for accurate downrange readings would undoubtedly be easier if the instrument had an optical sighting mechanism or laser, and if the instrument somehow displayed signal-to-noise ratio as a function of range for each shot detected. (This is available in the Track files, but it is not displayed.)

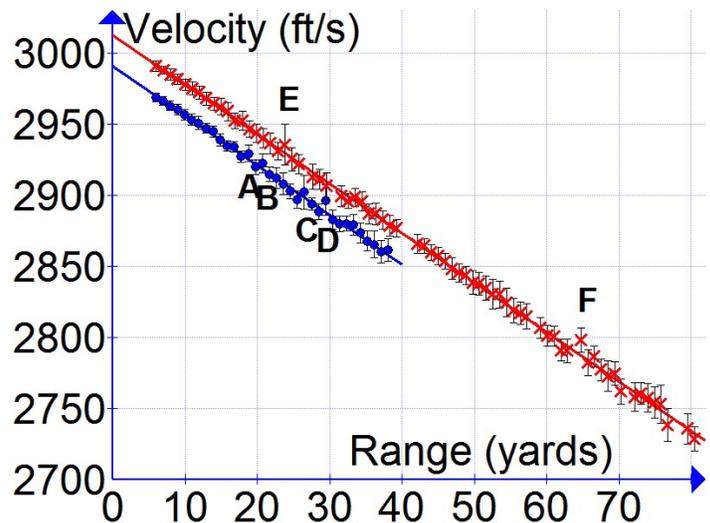

*Figure 2: Velocity vs. distance graphs from two Track files for the 22 caliber Hornady 55 grain VMAX bullet. The lower data set (blue circles) came from an earlier shot. The data set with the red Xs came from a later shot with presumably better overlap with the radar beam, because readings are available out to 10 yards beyond the maximum user range setting. Error bars are simply estimated as 100/SNR, where SNR is the radar signal-to-noise reported in the Track file.*

*Analysis Methods*

Figure 2 shows a graph of the raw data from two Track files downloaded from the instrument. The blue circles are the raw data from an early shot with imperfect alignment of the radar beam and the projectile path. This is clear, because the radar stopped reporting projectile velocity beyond 40 yards. Close inspection shows four data points (above the A, B, C, and D in the figure) where the radar reported a higher projectile velocity than the one reported for the next shortest range. This is a physical impossibility (bullets do not gain velocity in free flight), so it must represent measurement error.

# Accurate Measurements of Free Flight Drag Coefficients with Amateur Doppler Radar

These occurrences seem coincident with lower SNR, as shown by the larger error bars in the figure. (Error bars are estimated as 100/SNR). The pattern also holds for the velocities just below the E and F in the later shot shown in the graph (red Xs).

One analysis method (labeled "Report" hereafter) computes drag coefficients and ballistic coefficients using V0 and the velocity at a further distance from the Report file downloaded from the instrument. The available velocities in this file are determined from distances of interest pre-selected by the user. It is unclear how the unit determines these velocities, but some sort of regression seems to be used. Once a near and far velocity are available, drag coefficients are computed with the method of Courtney et al. (2014 and 2015) and ballistic coefficients are computed by entering the data into the JBM ballistic coefficient online calculator.

The second analysis method focuses on the slope of a best fit line to the data for velocity vs. distance downloaded in the Track file record of each shot. Though other methods are possible using the slope directly, it was convenient to use the slope to compute a second "far" velocity and then proceed with computing drag and ballistic coefficients in the same manner as before. Drag and ballistic coefficients computed from the raw Track file data will be labeled "Track" hereafter in this paper.

The least squares fitting process provides an uncertainty estimate in the slope, a unique feature to computing the drag and ballistic coefficients from the Track file data. Since a given relative error in the slope propagates nearly directly to the uncertainty in the drag and ballistic coefficients, this method allows one to estimate uncertainty in the drag and ballistic coefficients of each single shot. For example, in Figure 2, instrumental errors in the lower data set (blue circles) yield an uncertainty of 2.1% in the resulting best fit slope. This propagates to approximately a 2.1% relative uncertainty in the resulting drag coefficient and ballistic coefficient of that shot. In contrast, the errors in the upper data set (red Xs) result in a 0.5% uncertainty in the slope of the best fit line to that data set, propagating to approximately a 0.5% uncertainty in the resulting drag coefficient and ballistic coefficient of that shot.

After obtaining drag and ballistic coefficients from a number of shots for the same projectile, one can compute the mean, standard deviation, and standard error of the mean (SEM). As expected, projectiles with relatively small uncertainties on each shot have smaller standard deviations and SEMs. It is less clear whether using the raw Track data or the instrument generated Report data results in smaller SEMs for the resulting drag and ballistic coefficients.

In practice, there is some flexibility choosing distances over which to compute drag and ballistic coefficients. Understanding the trade offs between computing these over a shorter distance (closer to an instantaneous coefficient at a single velocity) and over a longer distance (more of an average coefficient over the range of velocities represented in the interval), it is convenient to compute drag coefficients over the first 20-25 yards in most cases. If the SNR is good (and it often is) over 50-100 yards, the drag and ballistic coefficients can also be computed over the longer distance.

*Sample size*

If measurement errors are random, the uncertainty in the mean drag coefficient can be reduced by increasing the sample size. The standard error of the mean (SEM) generally decreases as the square root of the sample size increases, so increasing the sample size by a factor of 4 can decrease the uncertainty in the mean by a factor of two. If a sample size of five is dominated by random errors and yields a SEM of 2%, then it is expected that a sample size of 20 would reduce the SEM to close to 1%.

One can usually use such a "brute force" approach to reducing uncertainties by increasing sample sizes. However, it is preferable to reduce instrumental and other experimental uncertainties first. In this study, a range of sample sizes was used. In part this was an intentional effort to demonstrate the accuracy potential of different sample sizes. Additionally, this was an unintentional consequence of the instrument sometimes failing to trigger or imperfect alignment leading to failures to capture data for every shot fired.

*Projectiles*

To assess measurement and accuracy capabilities of the instrument, a range of projectiles and velocities were used. In some cases, different powder charges were used to determine drag at different velocities. Full power rifle loads use standard reloading techniques from published sources and reloading manuals. Reduced velocity loads use H4895 and Sr4759 powders for .308 inch diameter bullets and Alliant Blue Dot Powder for .224 inch diameter

# Accurate Measurements of Free Flight Drag Coefficients with Amateur Doppler Radar

bullets. A variety of rifles and pistols were used in the appropriate calibers.

## Results: 30 Caliber Rifle Bullets (M1.2 to M2.8)

| Bullet | Wt gr | V ft/s | Cd Track | SEM | Cd Report | SEM | n |
|---|---|---|---|---|---|---|---|
| Sp FP | 150 | 1385 | 0.494 | 0.3% | 0.501 | 1.0% | 3 |
| DGHP | 130 | 1506 | 0.784 | 0.7% | 0.789 | 1.0% | 3 |
| FBBT | 155.5 | 1387 | 0.362 | 0.8% | 0.377 | 0.6% | 6 |
| Sp FB | 200 | 1172 | 0.368 | 0.9% | 0.390 | 2.0% | 21 |
| NP | 220 | 1306 | 0.429 | 0.6% | 0.436 | 1.2% | 5 |
| NP | 220 | 1206 | 0.397 | 1.6% | 0.412 | 1.7% | 5 |
| NET | 150 | 1257 | 0.496 | 0.9% | 0.507 | 0.6% | 12 |
| AMAX | 155 | 2878 | 0.283 | 1.5% | 0.287 | 3.6% | 4 |
| AMAX | 155 | 2544 | 0.316 | 1.8% | 0.332 | 2.0% | 10 |
| TMK | 175 | 2712 | 0.263 | 3.2% | 0.259 | 1.7% | 11 |
| ELD | 208 | 2888 | 0.242 | 1.3% | 0.244 | 1.3% | 12 |
| ELD | 208 | 1353 | 0.356 | 0.9% | 0.378 | 0.7% | 5 |
| HSP | 110 | 1758 | 0.456 | 0.5% | 0.472 | 0.8% | 12 |
| VMAX | 110 | 1745 | 0.433 | 0.4% | 0.451 | 0.8% | 20 |
| NBT | 125 | 1677 | 0.435 | 0.3% | 0.435 | 0.6% | 6 |
| TTSXBT | 168 | 2709 | 0.313 | 1.1% | 0.310 | 1.1% | 14 |
| SST | 125 | 2893 | 0.309 | 0.7% | 0.314 | 1.4% | 10 |

*Table 1: Results of drag coefficient (Cd) measurements for a number of 7.62mm projectiles (0.308 inch diameter). Abbreviations: SEM (standard error of the mean), n (sample size), Wt (weight), gr (grains), V (near velocity), Sp FP (Speer flat point), Sp FB (Speer flat base), DGHP (brass dangerous game hollow point, Cutting Edge Bullets), FBBT (Fullbore boat tail, Berger Bullets), NP (Nosler Partition), NET (Nosler E-Tip), AMAX (Hornady A-MAX), TMK (Sierra Tipped MatchKing), ELD (Hornady ELD), HSP (Hornady Spire Point), VMAX (Hornady V-MAX), NBT (Nosler Ballistic Tip), TTSXBT (Barnes Tipped Triple-Shock Boat Tail), SST (Hornady Super Shock Tip).*

Results of drag coefficient determination for a number of 0.308 inch diameter projectiles are shown in Table 1. In most cases, Cd values determined from the Track files (raw data) and from the Report files are within the uncertainties (SEM) of each other. In 11 of 17 cases, the Track method yielded smaller uncertainties; whereas, in 4 of 17 cases, the Report method yielded smaller uncertainties.

In 11 of 17 cases, the Track method yields uncertainties under 1%, even though sample sizes tend to be reasonably small. The Report method only yields uncertainties less than 1% in 6 of 17 cases.

The boat tail bullets (DGHP, FBBT, NET, AMAX, TMK, ELD, NBT, and TSXBT) tended to have larger uncertainties than the flat base bullets, though a number of boat tail bullets had uncertainties below 1% (FBBT, NET, ELD at 1353 ft/s, and NBT). In most cases, the uncertainties could probably be reduced to less than 1% with greater care in instrument alignment and/or increased sample size.

The projectiles represent a wide range of drag coefficients, from a high of 0.784 for the 130 grain DGHP which has a large open hollow point to a low of 0.242 for the 208 grain ELD at 2888 ft/s with a plastic tip, a boat tail, and a shape that is optimized to reduce drag. At 0.496, the drag coefficient of the 150 grain NET is uncharacteristically high for a pointed boat tail bullet. This is likely due in part to transonic drag rise, but Halloran et al. (2012) also noted a much higher drag than expected for this bullet at M2.5, ascribing it to a pronounced shoulder between the plastic tip and copper portion at the front of the projectile.

The velocity dependence of drag coefficients is easily seen in the three cases where data is available for two different near velocities (155 AMAX, 208 ELD, and 220 NP).

Drag coefficients of several of these same projectiles have also been measured by Litz (2009a) at comparable velocities using an acoustic method. Precise comparison is hindered by Litz reporting drag coefficients in graphs without giving uncertainties, but inspection of his graphs suggests agreement for the 155 AMAX and the 168 TTSXBT. Litz does not present data for the drag of the 155.5 FBBT below M1.5, but the extrapolation of the G7 drag model fit to his data agrees with the measurement reported here. The Cd for the 125 NBT reported here is much higher than that measured by Litz. The discrepancy of measuring much higher drag for this bullet from BTG Research test rifles than Litz measures with his test rifles is consistent and has previously been discussed. (Courtney and Courtney, 2007; Litz 2009b, Halloran et al., 2012).

*Detecting a tumbling projectile*

Even though Lehigh Defense recommends a 1 in 8" minimum twist rate for their 200 grain solid copper Maximum Expansion (ME) bullet, in the present study it was fired it with a reduced load in a 1 in 10" barrel. It was expected it might tumble, and it did as evidenced by keyholes (sideways bullet holes) in the target. Tumbling provided an opportunity to

# Accurate Measurements of Free Flight Drag Coefficients with Amateur Doppler Radar

see whether the LabRadar would provide a signature of tumbling bullets. A graph of velocity vs. distance (raw data) is shown in Figure 3, along with a best fit polynomial.

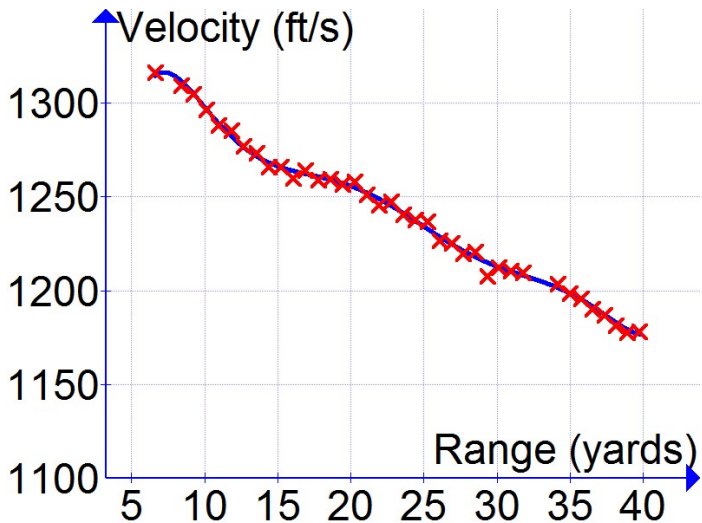

Figure 3: Raw velocity vs. distance data for tumbling 200 grain Lehigh Defense ME bullet along with best fit polynomial.

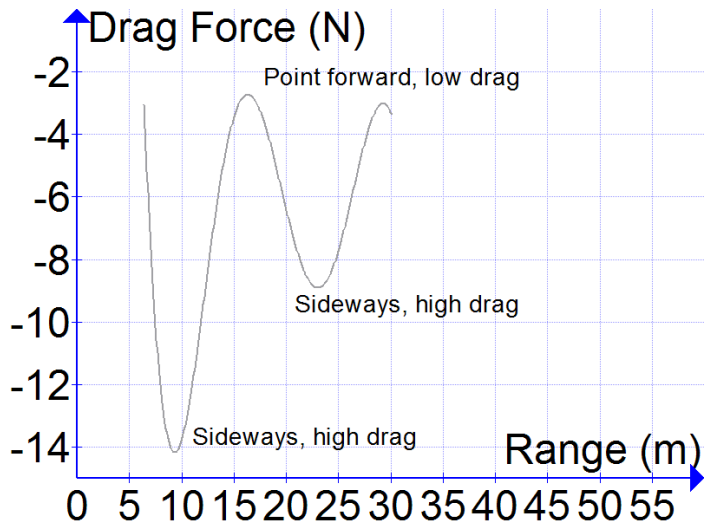

Figure 4: Drag force vs. projectile range for tumbling Lehigh Defense 200 ME fired at 1360 fps.

Unlike the case of bullets flying point forward which have a nearly constant (or slowly varying) rate of change, this bullet alternates between losing velocity slowly when it is point (or base) forward and quickly when it is moving sideways or at a large yaw angle relative to its direction of travel. Figure 4 shows a graph of the retarding force vs. range estimated by further analysis. In the present example, tum-bling was detected both in the paper target at 80 yards and in the Doppler radar velocities.

However, there are many cases in ballistics where tumbling leads to such large inaccuracies that there are no holes in the paper target (often further downrange). The LabRadar can provide a definitive signature of tumbling where one might otherwise be guessing regarding the reason for missing the paper target completely. Though not shown in the graphs, the SNR also oscillates, being larger at distances when the bullet is sideways and smaller when it is pointing more directly toward or away from the radar presenting a smaller area for reflection.

**Results: 22 Caliber Rifle Bullets (M1.6 to M3.0)**

| Bullet | Wt gr | V ft/s | Cd Track | SEM | Cd Report | SEM | n |
|---|---|---|---|---|---|---|---|
| TTSXBT | 55 | 2815 | 0.456 | 3.4% | 0.479 | 2.2% | 5 |
| VMAX | 53 | 2804 | 0.313 | 1.5% | 0.315 | 2.0% | 3 |
| BFB | 62 | 3075 | 0.358 * | 2.0% | 0.368 | NA | 1 |
| BFB | 62 | 2645 | 0.393 | 1.0% | 0.394 | 1.1% | 5 |
| BFB | 62 | 2479 | 0.410 | 1.2% | 0.411 | 2.1% | 2 |
| BFB | 62 | 2371 | 0.388 | 1.2% | 0.393 | 1.8% | 4 |
| BFB | 62 | 2223 | 0.422 | 1.0% | 0.422 | 1.1% | 3 |
| BFB | 62 | 2074 | 0.419 | 0.3% | 0.420 | 0.7% | 6 |
| BFB | 62 | 1877 | 0.446 | 1.8% | 0.447 | 1.9% | 4 |
| BFB | 62 | 1724 | 0.457 | 0.9% | 0.466 | 0.9% | 9 |
| VMAX | 55 | 3243 | 0.302 | 0.6% | 0.309 | 0.6% | 16 |
| SMK | 80 | 2823 | 0.282 | 0.7% | 0.288 | 1.2% | 8 |
| SMK pt | 80 | 2859 | 0.244 | 2.4% | 0.254 | 2.4% | 8 |

Table 2: Results of drag coefficient (Cd) measurements for a number of 5.56mm projectiles (0.224 inch diameter). Common abbreviations same as Table 1. Additional abbreviations: BFB (Berger Flat Base), SMK (Sierra MatchKing, unpointed), SMK pt (factory pointed Sierra MatchKing).*This is uncertainty from least squares fit to raw data rather than SEM.

Results of drag coefficient determinations for a number of 0.224 inch diameter projectiles are shown in Table 2. In most cases, the Cd values determined from the Track files (raw data) and from the Report files are within the uncertainties (SEM) of each other. In 8 of 12 cases, the Track method produces smaller uncertainties; whereas, in 1 of 12 cases the Report method yields smaller uncertainties.

In 10 of 12 cases, the Track method yields uncertainties at or under 2%, even though sample sizes tend to be reasonably small. Recall that with the .0.308 inch diameter bullets, the Track method

# Accurate Measurements of Free Flight Drag Coefficients with Amateur Doppler Radar

yielded uncertainties under 1% in the majority (11 of 17) of cases; the uncertainty is only equal to or less than 1% in 6 of 12 cases with the smaller diameter bullets. This is likely attributable to the smaller radar SNR with the smaller diameter bullets, but it may be reduced with more careful alignment procedures and/or larger sample sizes.

Results of loading a range of powder charges to map out Cd vs. velocity over a range of velocities for the 62 grain BFB are shown in Figure 5. There is no technical limitation to reducing the powder charges (using Blue Dot powder, Courtney and Miller 2012b) to measure drag coefficients down to M1.0, thus mapping the drag coefficients throughout the supersonic range. This bullet demonstrates the usual trend of decreasing Cd from M1.5 toward M3.0.

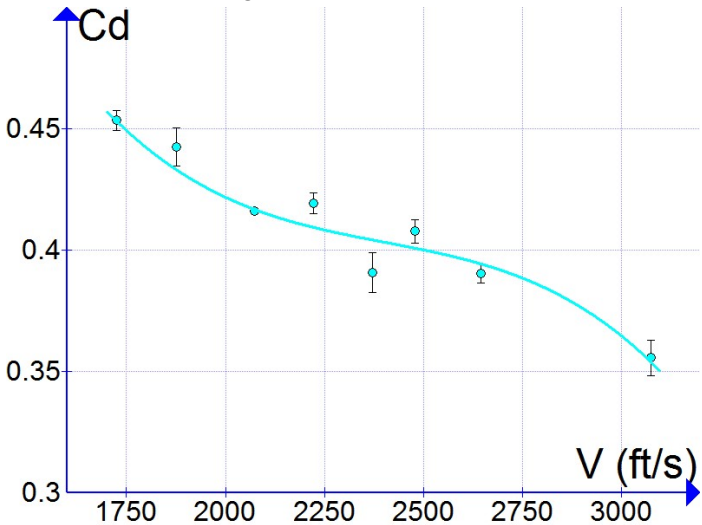

*Figure 5: Drag coefficient (Cd) vs. near velocity over a range of velocities for the 62 grain BFB.*

The lack of a strict monotonic decrease in drag suggests something unusual may be happening or that using the SEM to estimate the experimental uncertainties may not be accurate for such small sample sizes (2-9). Prior work (Courtney et al., 2014) has had better results using sample sizes of 10-20 shots at each velocity for mapping out Cd vs. velocity. In the present study, 10 rounds were prepared with each powder charge, but shots were missed, because this data was taken while details of instrumentation triggering and alignment were being fine tuned. Unfortunately, Berger Bullets has discontinued this bullet, so the experiment shown in Figure 5 cannot be repeated with a larger sample size.

The last two rows of Table 2 show results of pointed and unpointed 80 grain Sierra MatchKings. There is a small difference in near velocities, but most of the difference in drag coefficients is likely attributable to the pointing. Sierra began pointing their 90 and 80 grain SMKs in the past few years, because their internal results showed significant reductions in drag. A drag reduction of 13% may seem on the large side for only pointing the tip of the projectile; however, Sierra's own measurements report a 12% drag reduction due to pointing their 90 grain SMK. (Sierra has not yet published new drag measurements for the 80 grain SMK, but a private communication suggests a decrease in drag of 10% by pointing.)

There are theoretical approaches (McCoy, 1981) that partition contributions to the drag coefficient into nose drag, base, drag, skin friction, etc. The ability of Doppler radar to detect drag decrease due to pointing suggests the possibility of carefully studying the contributions of other projectile modifications also. For example, one might measure the drag coefficients of (otherwise identical) machined brass projectiles with varying boat tail angles (or no boat tail at all).

**Results: 9mm Pistol Bullets (M1.1 to M0.9)**

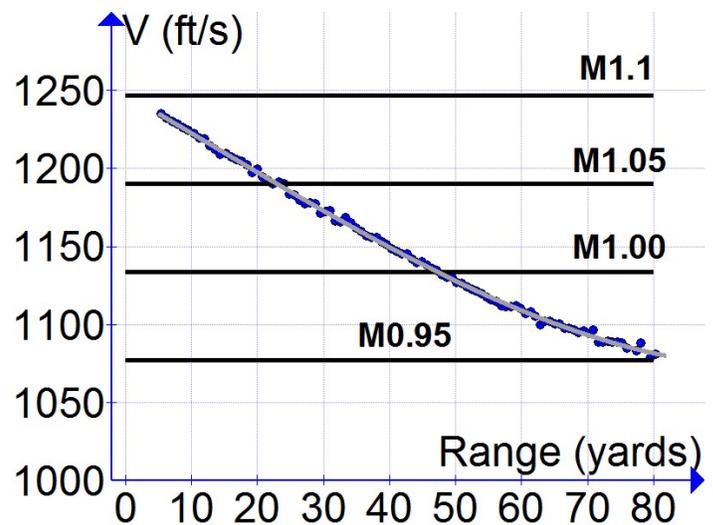

*Figure 6: Velocity vs. distance for a pistol bullet along with best fit cubic polynomial and horizontal lines corresponding to Mach numbers.*

A typical Track file showing the raw data of velocity vs. distance is shown in Figure 6 for pistol bullets with muzzle velocity near 1250 ft/s. Note that due to high drag and relatively light mass, combined with the sonic transition, a clear change in the local rate of velocity loss is apparent over the flight. This sug-

# Accurate Measurements of Free Flight Drag Coefficients with Amateur Doppler Radar

gests the possibility of using the raw data files to determine how drag is changing in each Mach band. (See Section 4.5 of McDonald and Algren, 1995.)

This was done for the 115 grain Hornady XTP bullet in 9mm, and results are shown in Table 3. The top edge of the Mach bands is shown in the velocity column. Cd changes little from M1.10 to M1.05, but then there is a steep drop from the lowest supersonic band to the highest subsonic band with Cd dropping from 0.605 to 0.475. The accuracy is better than 1% in each case, suggesting that the LabRadar may be a useful tool for studying transonic drag changes in a variety of cases.

Table 3 also shows drag coefficients for several pistol bullets as determined in their first 25 yards of flight from the Report and Track files. As expected, the subsonic 147 grain WWB load has a much lower drag coefficient than the supersonic bullets.

| Bullet | Wt gr | V ft/s | Cd Track | SEM | Cd Report | SEM | n |
|---|---|---|---|---|---|---|---|
| XTP | 115 | 1252 | 0.616 | 0.3% | 0.611 | 0.5% | 28 |
| XTP | 115 | M1.10 | 0.609 | 0.7% | NA | NA | 20 |
| XTP | 115 | M1.05 | 0.605 | 0.5% | NA | NA | 20 |
| XTP | 115 | M1.00 | 0.475 | 0.7% | NA | NA | 20 |
| FMJ | 115 | 1176 | 0.604 | 1.4% | 0.603 | 1.0% | 11 |
| WWB | 147 | 969 | 0.256 | 1.4% | 0.264 | 1.5% | 10 |

Table 3: Results of drag coefficient (Cd) measurements for a number of 9mm projectiles (0.355 inch diameter). Common abbreviations same as Table 1 and 2. Additional abbreviations: M (Mach), XTP (Hornady XTP), FMJ (Winchester Full Metal Jacket NATO ball), WWB (Winchester White Box JHP – Jacketed Hollow Point), SXT (Winchester Ranger SXT JHP law enforcement duty load).

## Results: 4.4 mm bbs (M0.3 to M0.5)

| Bullet | Wt gr | V ft/s | Cd Track | SEM | Cd Report | SEM | n |
|---|---|---|---|---|---|---|---|
| bb | 5.44 | 368 | 0.451 | 0.9% | 0.432 | 1.3% | 10 |
| bb | 5.44 | 547 | 0.480 | 1.4% | 0.473 | 0.7% | 10 |

Table 4: Results of drag coefficient (Cd) measurements for 0.177 inch diameter steel bbs (Avanti). Common abbreviations same as Table 1 and 2.

Results of drag coefficients for 0.177 inch diameter steel spheres (bbs) are shown in Table 4. The accuracy is not as good as obtained with the dual chronograph method (Figure 2 of Courtney et al., 2015) where the drag coefficient was mapped from 90 to 180 m/s with uncertainties below 1%. Since these were the same bbs fired from the same bb gun, it is reasonable to presume that the shot to shot variations in drag were also small in the LabRadar experiment and that the slightly larger uncertainties are attributable to the instrument rather than shot to shot variations in the projectiles. The spherical surface of a 0.177 inch diameter bb is a poor radar reflector, and signal-to-noise ratios were low, even over the relatively short 30 foot distance of this experiment.

For educational labs, it might be desirable to employ the LabRadar to measure drag of subsonic projectiles that are safer than firearms. Since the small size of 0.177 inch diameter bbs limits measurement accuracy, it may be preferable to use larger diameter (6mm or 8mm, 0.24 to 0.32 inch) Airsoft pellets. We would recommend the higher quality (more perfect spheres) ceramic projectiles. Another option is the even larger diameter (0.68 inch) paintballs from an appropriate paintball marker; however, paintballs are soft, not nearly perfect spheres, require more cleanup, and may produce larger shot to shot variations in drag.

## Discussion

Strengths of using the LabRadar to measure drag in free flight include low expense, ease of use, accuracy, and usability over a wide range of Mach numbers and different projectiles. The main limitations are accuracy and alignment difficulty with smaller calibers, and limited accuracy and range with spherical projectiles and other projectiles without flat rear surfaces. Limited range may hinder measurement of ballistic coefficients appropriate for predicting long range trajectories, but this may be remedied by measuring drag over the entire velocity range of interest by using reduced loads.

There are a number of appealing possibilities for educational lab use, including using Airsoft and paintball projectiles, ability to measure drag in one experiment and then use measured drag to predict trajectory in a subsequent experiment, and the possibility of a detailed study of drag changes through the sonic transition. Many Airsoft, paintball, and bb projectile launchers have a mechanism for adjusting muzzle velocity. This is convenient for studying drag over a range of velocities and to study velocity dependencies, if that is desired.

In more advanced aerodynamics, the availability of plastic sabots allows the study of drag at very high Mach numbers (M4.5) using

# Accurate Measurements of Free Flight Drag Coefficients with Amateur Doppler Radar

conventional firearms as the projectile launchers. Combined with lathe turned projectiles of different shapes, sabots may be used to study the effect of a variety of independent variables on drag at greater accuracy and lower cost than previously available.

The LabRadar should also empower the study of a number of practical matters in external ballistics of interest to military, law enforcement, and sporting applications. These include mapping of drag curves over a broad range of Mach numbers for more accurately computing long range (800-1200 yards) and very long range (1200-2500 yard) trajectories. Most very long range trajectories include a sonic transition even with the lowest drag projectiles currently available.

In the past few years, measuring drag accurately has taken on increasing importance in understanding both bullet stability as well as the damping of pitch and yaw for spin stabilized projectiles (Courtney and Miller, 2012b; Litz, 2009; Courtney et al., 2012; McDonald and Algren, 1995, 2003). Since effects on bullet stability and damping of pitch and yaw are easily apparent in the first 50-100 yards, the LabRadar seems particularly suited to improving precision of these studies with reduced costs and risks to equipment in front of the firing line. LabRadar can also provide an estimate for how far downrange bullets tumble and how many times a bullet tumbles in a given distance.

Bullet manufacturers have occasionally noted impacts on bullet drag with die wear and the replacing of dies and other equipment used in forming projectiles. The LabRadar may provide a convenient and inexpensive means to check for drag changes in the first 50-100 yards without more expensive and cumbersome methods for measuring drag effects over longer ranges. The LabRadar may also provide rapid feedback on design changes or modifications, not only in the projectiles but also in barrels (Bohnenkamp et al., 2011). Use of the LabRadar on the firing line of long range matches may provide a Physics based approach to diagnosing dropped points.

Those who use the LabRadar to compute drag coefficients face the practical choice of using the raw data from the Track files or the displayed velocities which also appear in the Report files. Beginning with the raw data is more accurate most of the time; however, performing a least squares fit for each shot may be time consuming. In most cases, Report data may be used to save time without sacrificing more than a factor of two in accuracy (and in some cases, it is more accurate). Depending on the accuracy needs of the project, it is simple enough to compute drag coefficients quickly with the Report data, turning to the raw data from the Track files if greater accuracy is desired after estimating the uncertainty in the Report results by computing the SEM.

**Appendix: Ballistic Coefficients**

The ballistic coefficients for the .308 inch diameter rifle bullets used in this study are shown in Table A1. Since the G7 drag model is more appropriate for boat tail bullets, it is not shown for flat base bullets. The uncertainty (SEM) in the G1 BC is 1.0% or less for 10 of the 17 cases tested. As expected, the G7 BC shows much less change with different velocities than the G1 BC for boat tail bullets tested at more than one velocity (the 208 ELD and the 155 AMAX).

The G7 BC of the 208 ELD of 0.348 agrees well with the manufacturer's value of 0.335 and suggests this bullet would have high retained energy and low wind drift at extended ranges, especially since the G7 BC only drops to 0.337 at 1353 ft/s. Table A1 highlights how simple it is to measure rifle bullet BCs below M1.5 using the LabRadar with light loads. Sierra is the only bullet manufacturer to publish measured BCs over the whole supersonic range, but they only publish G1 BCs for most bullets. Berger Bullets chief ballistics expert, Bryan Litz, has published BCs for many Berger bullets that are purportedly accurate from M1.0 to M1.5 (Litz 2009a), but the reported values may be extrapolated from values measured at higher velocities, since the published graphs (Litz 2009a) contain very few measured drag values below M1.5.

The ballistic coefficients for the .224 inch diameter rifle bullets used in this study are shown in Table A2. The boat tail bullets (80 grain SMK, 53 grain VMAX, and 55 grain TTSXBT) tend to have larger uncertainties than the flat base bullets, especially considering the larger sample size. This is likely because of the reduced SNR from the Doppler radar reflection of the smaller effective base area. This uncertainty may be reduced by more careful alignment of the Doppler radar and/or larger sample sizes.

# Accurate Measurements of Free Flight Drag Coefficients with Amateur Doppler Radar

| Bullet | Wt gr | V ft/s | n | G1BC Track | SEM | G7BC Track | SEM |
|---|---|---|---|---|---|---|---|
| Sp FP | 150 | 1385 | 3 | 0.300 | 0.2% | NA | NA |
| DGHP | 130 | 1506 | 3 | 0.164 | 0.6% | NA | NA |
| FBBT | 155.5 | 1387 | 6 | 0.415 | 0.8% | 0.251 | 0.7% |
| Sp FB | 200 | 1172 | 21 | 0.410 | 0.8% | NA | NA |
| NP | 220 | 1306 | 5 | 0.469 | 0.9% | NA | NA |
| NP | 220 | 1206 | 5 | 0.445 | 1.6% | NA | NA |
| NET | 150 | 1257 | 12 | 0.259 | 1.1% | 0.184 | 0.8% |
| AMAX | 155 | 2878 | 4 | 0.446 | 1.5% | 0.221 | 3.7% |
| AMAX | 155 | 2544 | 10 | 0.419 | 1.6% | 0.211 | 1.6% |
| TMK | 175 | 2712 | 11 | 0.559 | 3.4% | 0.281 | 3.4% |
| ELD | 208 | 2888 | 12 | 0.697 | 1.4% | 0.348 | 1.4% |
| ELD | 208 | 1353 | 5 | 0.550 | 1.0% | 0.337 | 0.9% |
| HSP | 110 | 1758 | 12 | 0.234 | 0.5% | NA | NA |
| VMAX | 110 | 1745 | 20 | 0.247 | 0.3% | 0.130 | 0.3% |
| NBT | 125 | 1677 | 6 | 0.286 | 0.3% | 0.154 | 0.1% |
| TTSXBT | 168 | 2709 | 14 | 0.448 | 1.1% | 0.225 | 1.0% |

Table A1: Results of BC measurements for 7.62mm projectiles (0.308 inch diameter). Abbreviations are the same as in Table 1. G1BC (Ballistic coefficient with G1 drag model), G7BC (Ballistic coefficient with G7 drag model).

| Bullet | Wt gr | V ft/s | n | G1BC Track | SEM | G7BC Track | SEM |
|---|---|---|---|---|---|---|---|
| TTSXBT | 55 | 2815 | 5 | 0.189 | 3.2% | 0.095 | 3.2% |
| VMAX | 53 | 2804 | 3 | 0.264 | 1.4% | 0.132 | 1.5% |
| BFB | 62 | 3075 | 1 | 0.261 | 2.0% | NA | NA |
| BFB | 62 | 2645 | 5 | 0.252 | 0.9% | NA | NA |
| BFB | 62 | 2479 | 2 | 0.249 | 1.0% | NA | NA |
| BFB | 62 | 2371 | 4 | 0.265 | 1.9% | NA | NA |
| BFB | 62 | 2223 | 3 | 0.254 | 0.9% | NA | NA |
| BFB | 62 | 2074 | 6 | 0.263 | 0.2% | NA | NA |
| BFB | 62 | 1877 | 4 | 0.257 | 1.6% | NA | NA |
| BFB | 62 | 1724 | 9 | 0.255 | 0.8% | NA | NA |
| VMAX | 55 | 3243 | 16 | 0.266 | 0.6% | NA | NA |
| SMK | 80 | 2823 | 8 | 0.440 | 0.6% | 0.220 | 0.6% |
| SMK pt | 80 | 2859 | 8 | 0.508 | 2.5% | 0.254 | 2.5% |

Table A2: Results of BC measurements for 5.56mm projectiles (0.224 inch diameter). Abbreviations are the same as in Table 1, 2, and A1.

The 53 grain VMAX has a measured G1BC (0.264) significantly lower than the value claimed by Hornady (0.290). This might have been expected, since the value advertised by Hornady is much larger than any other 5.56mm bullet at or under 55 grains. At the same time, the 55 grain VMAX has a measured G1BC (0.266) slightly higher than advertised by Hornady (0.255). This may be due to the velocity at which it was measured (3243 ft/s) in the present study. Since the 53 grain VMAX has a short boat tail and the 55 grain VMAX does not, these results suggest that short boat tails may contribute much for drag reduction of 5.56 mm bullets.

The large increase in BC from pointing was unexpected for the 80 grain SMK, but it is consistent with the measured increase in BC published by Sierra for their similar 90 grain bullet in the same caliber. It would be of interest to study the BC impact of pointing over a larger range of velocities.

Ballistic coefficients for the 9mm pistol bullets used in this study are shown in Table A3. Hornady advertises a BC of 0.129 for their 115 grain XTP, and while that is what we measured from M1.0 to M0.95, the drag is rapidly changing through the transonic region. If accurate ballistic coefficients are needed or if long range trajectories need to be accurately computed, it may be advisable to measure drag more directly throughout the velocity region of interest.

| Bullet | Wt gr | V ft/s | n | G1BC Track | SEM |
|---|---|---|---|---|---|
| XTP | 115 | 1252 | 28 | 0.118 | 0.3% |
| XTP | 115 | M1.10 | 20 | 0.124 | 0.7% |
| XTP | 115 | M1.05 | 20 | 0.114 | 0.4% |
| XTP | 115 | M1.00 | 20 | 0.129 | 0.7% |
| FMJ | 115 | 1176 | 11 | 0.112 | 1.4% |
| WWB | 147 | 969 | 10 | 0.187 | 1.4% |
| SXT | 127 | 1234 | 13 | 0.151 | 1.3% |

Table A3: Results of BC measurements for 9mm projectiles (0.355 inch diameter). Abbreviations are the same as in Table 1, 2, 3, A1, and A2. Velocities designated by Mach numbers refer to the near velocity (top end of the band) and extend M0.05 lower.

Winchester's published trajectory information corresponds to a G1BC of 0.206 for the 147 grain WWB load. Our measured BC of the same factory load is 0.187. Likewise, Winchester's published trajectory corresponds to a G1BC of 0.146 for their 115 grain FMJ load; whereas, our measured BC is much lower at 0.112. Further, Winchester's published trajectory for their 127 grain SXT load corresponds with a BC of 0.164; whereas, the LabRadar measurements suggest a BC of 0.151.

# Accurate Measurements of Free Flight Drag Coefficients with Amateur Doppler Radar

These discrepancies suggest that if accurate trajectories are needed, a more accurately measured value should be used.

Earlier work (Courtney and Courtney, 2007; Halloran et al., 2012; Bohnenkamp et al., 2012) has shown that BCs of many rifle bullets are often reported to be higher by their manufacturers than when measured independently. The present study suggests BC exaggerations may be fairly common among pistol bullets also. Because pistol bullets tend to be used at shorter ranges, accurate drag measurements receive less attention. However, the present study suggests the need for better care in BC determination if selecting bullets for longer range applications or reconstructing shooting events over longer distances.

## Acknowledgments

This work was funded by BTG Research (www.btgresearch.org) and the United States Air Force. Elijah Courtney assisted with the data collection. Elijah Courtney and Amy Courtney provided valuable input into the experimental design, data analysis approach, and draft manuscript. Don Miller inspired the idea of using reduced charges to map drag coefficients vs. velocity. Bryan Litz inspired the idea of to use raw velocity data from the Track files rather than the displayed velocities from the instrument panel and Report files. The authors appreciate valuable feedback from several external reviewers which has been incorporated into revisions of the manuscript.

# Accurate Measurements of Free Flight Drag Coefficients with Amateur Doppler Radar

Miller, Don. How Good Are Simple Rules for Estimating Rifle Twist. Precision Shooting. June 2009, pp. 48-52.

## Author Addresses

ELYA COURTNEY
BTG Research
9574 Simon Lebleu Road
Lake Charles, LA, 70607

COLLIN MORRIS
Georgia Institute of Technology
Atlanta, GA, 30332

MICHAEL COURTNEY
BTG Research
9574 Simon Lebleu Road
Lake Charles, LA, 70607
Michael_Courtney@alum.mit.edu

## About the Authors

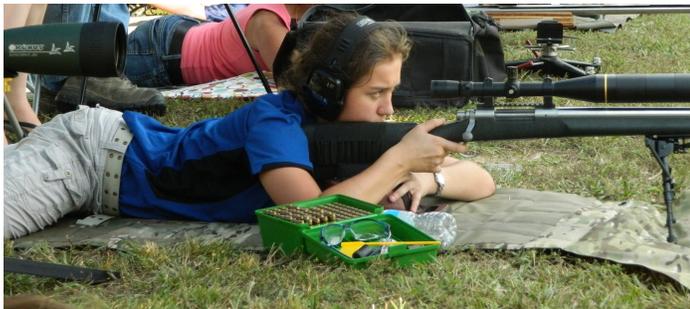

Elya Courtney is shown above shooting in a 1000 yard F-Class match. Elya has an NRA Expert rating at Mid-Range and shot the highest centerfire rifle scores in the NRA Women on Target national match in both 2014 and 2015, including a 200-16x in 2015. Elya won first place in the Louisiana State science fair three years in a row with projects in Physics, Chemistry, and Mathematics, including two projects in ballistics. She is attending a top 30 university majoring in Chemistry on a full academic scholarship.

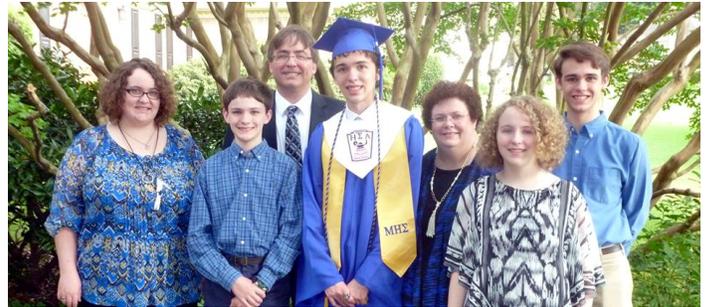

Collin Morris (above center) comes from a large family of Georgia Tech Alumni with a love of engineering, science, and medicine. He is a Mechanical Engineering Major at Georgia Tech with a particular interest in aerodynamics. He was invited to join the BTG Research team on this project after doing an outstanding job recognizing the limitations of using high speed video to determine drag coefficients of free flight projectiles.

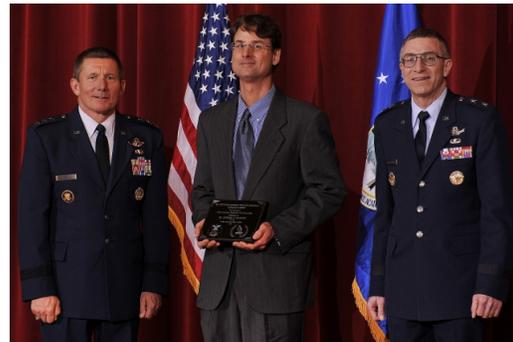

Michael Courtney (above center) is shown above receiving a research award from two Air Force Generals. He earned a PhD in Physics from MIT in 1995 and worked as an RF engineer before founding BTG Research in 2001 to study ballistics with applications in military and law enforcement. Dr. Courtney has a published dozens of scholarly papers in ballistics, is a founding member of the International Ballistics Society where he serves on their Publications Committee peer reviewing numerous abstracts and papers each year. Dr. Courtney's most notable works in external ballistics include two papers with Don Miller revising the Miller Stability Formula for plastic-tipped bullets, further revising the Miller Stability Formula for open-tipped match bullets, and experimentally verifying the improved stability formulas and the oft-assumed independence of bullet drag coefficients on air density. He has also co-authored papers in internal ballistics quantifying the friction effects of bullet coatings and documenting the performance problems of lead free primers.